\documentclass[twocolumn]{autart} \usepackage[dvipsnames]{xcolor}
\usepackage{color, soul}
\usepackage{mathtools}
\usepackage[pagewise]{lineno}
\usepackage{subcaption}
\usepackage{harvard}
\usepackage{float}
\usepackage{amsmath}
\usepackage{amssymb}
\usepackage{mathrsfs}
\usepackage[mathscr]{euscript}
\usepackage{arydshln}
\usepackage{multirow}
\usepackage{multicol}
\usepackage{tikz}
\usetikzlibrary{shapes,arrows}
\usepackage{verbatim}
\usepackage{lineno,hyperref}
\usepackage{xcolor}
\journal{Automatica}

\begin{document}

\begin{frontmatter}

\title{On the prescribed-time attractivity and frozen-time eigenvalues of linear time-varying systems\thanksref{footnoteinfo}} 

\thanks[footnoteinfo]{This paper was not presented at any IFAC 
meeting. Corresponding author A.~Shakouri. Tel. +98-935-2222206.}

\author{Amir Shakouri}\ead{a\_shakouri@outlook.com}

\address{Department of Aerospace Engineering, Sharif University of Technology, Tehran 14588-89694, Iran.}


\begin{abstract}
A system is called prescribed-time attractive if its solution converges at an arbitrary user-defined finite time. In this note, necessary and sufficient conditions are developed for the prescribed-time attractivity of linear time-varying (LTV) systems. It is proved that the frozen-time eigenvalues of a prescribed-time attractive LTV system have negative real parts when the time is sufficiently close to the convergence moment. This result shows that the ubiquitous singularity problem of prescribed-time attractive LTV systems can be avoided without instability effects by switching to the corresponding frozen-time system at an appropriate time. Consequently, it is proved that the time-varying state-feedback gain of a prescribed-time controller, designed for an unknown linear time-invariant system, approaches the set of stabilizing constant state-feedback gains.
\end{abstract}

\end{frontmatter}

\section{Introduction}
\label{sec:I}
Prescribed-time systems are characterized by their ability to converge at an arbitrary finite time and their robustness to unstable non-vanishing disturbances. In terms of the user's knowledge about the settling time, in finite-time schemes \cite{bhat2000finite,engel2002continuous}, it is only known that the system non-asymptotically converges at a finite time that is generally a function of the initial conditions. Fixed-time schemes provide an upper bound for the settling time, independent of initial conditions \cite{polyakov2011nonlinear,lopez2018finite}. However, in a prescribed-time scheme, which is essentially time-varying, the settling time is commanded to the system, which means that the user is not only aware of the convergence moment but can arbitrarily specify it just by changing a parameter. Over the last few years, prescribed-time controllers and observers have been developed in the literature for different types of systems, which the reader can refer to \citeasnoun{song2017time}, \citeasnoun{holloway2019prescribed}, \citeasnoun{holloway2019prescribed2}, \citeasnoun{krishnamurthy2020dynamic}, \citeasnoun{zhou2020finite}, \citeasnoun{zhou2020finite2}, \citeasnoun{ding2020nonsingular},  \citeasnoun{zhou2021prescribed}, \citeasnoun{shakouri2021framework}, and \citeasnoun{shakouri2021euler}. 

A prescribed-time controller (or observer) can be designed for unknown linear time-invariant (LTI) systems using time transformation mapping techniques \cite{holloway2019prescribed,holloway2019prescribed2,ding2020nonsingular,shakouri2021framework} or parametric Lyapunov equations \cite{zhou2020finite2,zhou2021prescribed} without a complete knowledge of the system eigenvalues. In this case, the closed-loop dynamics of an LTI system under a prescribed-time controller (or observer) is a linear time-varying (LTV) system. The stability analysis of LTV systems is considerably more difficult than the case of LTI systems \cite{blondel2012open,zhou2016asymptotic}. The eigenvalues of an LTV system at each time instant, called the frozen-time eigenvalues, can only be used as a measure of stability for slowly varying systems \cite{amato1993new}. Generally, the sign of the frozen-time eigenvalues cannot be used as a necessary or sufficient condition for the stability analysis of LTV systems\footnote{For example, a periodic LTV system with a constant positive frozen-time eigenvalue can be asymptotically stable \cite{wu1974note}.}. 

For an LTI system $\dot{x}=Fx+Gu$, a time-varying controller $u=K(t,\tau)x$ can be designed (independent of $F$ and $G$) such that the solutions of the closed-loop LTV system $\dot{x}=(F+GK(t,\tau))x\triangleq A(t,\tau)x$ converge to zero equilibrium at the commanded time $\tau>0$ such that $u=K(t,\tau)x$ is bounded \cite{zhou2020finite2}. In this case, although it is known that the closed-loop system $\dot{x}=A(t,\tau)x$ is prescribed-time attractive (PTA) at $t=\tau$, the frozen-time eigenvalues of $A(t,\tau)$ cannot be explicitly obtained if $F$ and $G$ are not completely available. In addition, the prescribed-time schemes suffer from a singularity at the convergence time. The source of this singularity is that, despite the boundedness of $u=K(t,\tau)x$, the time-dependent gain $K(t,\tau)$ approaches infinity as $t\rightarrow\tau$, which imposes problems in the practical implementation of these methods. As a solution, in \citeasnoun{zhou2020finite}, \citeasnoun{zhou2021prescribed}, and \citeasnoun{shakouri2021euler} a switching method is proposed by which the gains' increment is stopped at a sufficiently short time before the singularity happens, and afterward, the last calculated gains are used as constant values. In other words, speaking in terms of LTV systems, this strategy is based on switching from the PTA system to the corresponding frozen-time system (evaluated at the switching moment). This switching method is proposed in the literature as an intuitive solution and its viability has not been thoroughly analyzed. In order to prevent instability after switching, the frozen-time eigenvalues should have negative real parts at least close to the convergence time. When the stability of the closed-loop frozen-time system is guaranteed in a neighborhood of the equilibrium (independent of the open-loop LTI system), then the time-varying gain of the prescribed-time controller reaches the set of stabilizing constant state-feedback gains without relying on the system model. Therefore, the study of the frozen-time eigenvalues of PTA systems is motivated by preventing the divergence risk and retaining closed-loop stability after reaching a neighborhood of the equilibrium, especially when the open-loop LTI system matrices $F$ and $G$ are unknown.

In this note, we deal with general LTV systems. The results are not limited to prescribed-time schemes, and they are also valid for finite and fixed-time methods. It is proved in Theorem \ref{th:1} that the $p$-norm of the system matrix should approach infinity as a necessary condition for prescribed-time attractivity, i.e., every prescribed-time attractive LTV system is essentially singular at the convergence time. Theorem \ref{th:2} presents a necessary and sufficient prescribed-time attractivity condition in terms of the differential Lyapunov equation. Finally, Theorem \ref{th:3} discusses the frozen-time eigenvalues, which is the core result of this paper. An example is presented in Section \ref{sec:IV} to demonstrate some applications.  

\section{Preliminaries}
\label{sec:II}

\subsection{Notations}

Let $\mathbb{R}^{m\times n}$ denote the space of $m \times n$ real matrices, $\mathbb{R}^n$ denote the space of $n$-dimensional real vectors, and $\mathbb{N}$ denote the space of natural numbers. The $n$-dimensional identity matrix is denoted by $\mathbb{I}_n$. The $i$th entry of vector $v\in \mathbb{R}^n$ is referred to by $v_i$. Let $M_{ij}$ denote the entry of matrix $M$ on the $i$th row and $j$th column. The real part of the $i$th eigenvalue of matrix $M$ is shown by $\lambda_i(M)$. The maximum and minimum eigenvalues of a matrix $M$ are defined as $\lambda_{\mathrm{max}}(M)=\max_{i\in\mathbb{N}}\{\lambda_i(M)\}$ and $\lambda_{\mathrm{min}}(M)=\min_{i\in\mathbb{N}}\{\lambda_i(M)\}$, respectively. The $n\times n$ diagonal matrix with entries of vector $v\in\mathbb{R}^n$ as its diagonal elements is denoted by $\mathrm{diag}(v)$. The absolute value of scalars is shown by $|\cdot|$. Symbol $\|\cdot\|_p$ stands for the vector $p$-norm or the matrix $p$-norm (induced by vector $p$-norm). We use $\|\cdot\|$ instead of $\|\cdot\|_2$ for simplicity. The logarithmic norm of matrix $M\in\mathbb{R}^{n\times n}$ is defined as follows \cite{strom1975logarithmic}:
\begin{equation}
\label{eq:note-1}
\mu_p(M)=\lim_{h\rightarrow0^+}\frac{\|\mathbb{I}_n-hM\|_p-1}{h}
\end{equation}

\subsection{Problem Formulation}

Consider an LTV closed-loop system as follows:
\begin{equation}
\label{eq:1}
\dot{x}=A(t,\tau)x
\end{equation}
where $x\in\mathbb{R}^n$ is the state vector, $\tau>0$ is a user-defined constant, and $A(\cdot,\cdot):[0,\infty)\times(0,\infty)\rightarrow\mathbb{R}^{n\times n}$ is a matrix-valued function. 

\begin{defn}{\textbf{\textup{\cite{shakouri2021euler}}}
\label{def:1}
System \eqref{eq:1} is called \textit{(globally uniformly) prescribed-time attractive (PTA)} if for every $\tau>0$ (independent of initial state and time) its solution satisfies:
\begin{equation}
\label{eq:def:1-1}
\lim_{t\rightarrow\tau^-}\|x(t)\|=0
\end{equation}}
\end{defn}

The \textit{frozen-time system} corresponding to \eqref{eq:1} is the instantaneous LTI system evaluated at each time, which its eigenvalues are called \textit{frozen-time eigenvalues}. 

We say system \eqref{eq:1} is \textit{singular} if some entries of the system matrix approach infinity (or some denominators approach zero). A system that is not singular is called \textit{nonsingular}. A singular system can be stable in theory such that the state vector and its derivatives have finite values at all times. However, the singularity imposes problems in the numerical implementation of that system.  

It is impossible for a nonsingular LTV system (and all LTI systems) to be PTA or to show any finite escape time (see Corollary \ref{cor:1}--item 1). However, a nonsingular LTV system may reach an arbitrary nonzero error bound at a commanded settling time $\tau$ and retain that bound for $t\geq\tau$ \cite{ding2020nonsingular}. This type of nonsingular LTV systems possesses all practical advantages of PTA systems and is useful in the successfully implementing prescribed-time controllers and observers to (unknown) LTI systems. Therefore, we state the problem statements of this note as follows:

\begin{prob}
\label{prob:1}
Is it possible for a nonsingular LTV system expressed as $\dot{x}=A_s(t,\tau,\sigma)x$ to satisfy $\|x(t)\|\leq\sigma,\forall t\in[\tau,\infty)$, for every user-defined parameters $\sigma>0$ and $\tau>0$? 
\end{prob}

\begin{prob}
\label{prob:2}
Given that $u=K(t,\tau)x$ is a singular PTA controller for the (unknown) LTI system $\dot{x}=Fx+Gu$, find a nonsingular control law $u=K_s(t,\tau,\sigma)x$ such that $\|x(t)\|\leq\sigma,\forall t\in[\tau,\infty)$, for every user-defined parameters $\sigma>0$ and $\tau>0$.
\end{prob}

\section{Main Results}
\label{sec:III}

In this section, we present the main results of the paper in three theorems. The explicit answers to Problems \ref{prob:1} and \ref{prob:2} are given in Corollaries \ref{cor:2} and \ref{cor:3}, respectively. 

First, Consider the following lemma that is used in the proof of Theorem \ref{th:1}:

\begin{lem}\textbf{\textup{\cite{vrabel2019note}}}
\label{lem:1}
The solution of system \eqref{eq:1} satisfies the following condition:
\begin{equation}
\label{eq:lem:1-1}
\begin{split}
&\|x(0)\|_p\exp\left(-\int_{0}^t\mu_p[-A(s,\tau)]ds\right)\leq\|x(t)\|_p\\
&\leq\|x(0)\|_p\exp\left(\int_{0}^t\mu_p[A(s,\tau)]ds\right)
\end{split}
\end{equation}
\end{lem}

Given the upper and lower bounds of Lemma \ref{lem:1}, the following theorem can be directly concluded, which presents separate necessary and sufficient conditions for the prescribed-time attractivity of LTV systems.

\begin{thm}
\label{th:1}
System \eqref{eq:1} is PTA at $t=\tau$ if:
\begin{equation}
\label{eq:th:1-1}
\lim_{t\rightarrow\tau^-}\int_{0}^t\mu_p[A(s,\tau)]ds=-\infty
\end{equation}
In addition, if system \eqref{eq:1} is PTA at $t=\tau$ and $\lim_{t\rightarrow\tau-}|A_{i,j}(t,\tau)|$ exists or is infinity for all $i,j\in\mathbb{N}$, then:
\begin{equation}
\label{eq:th:1-2}
\lim_{t\rightarrow\tau^-}\mu_p[-A(s,\tau)]=\infty
\end{equation}
and thus:
\begin{equation}
\label{eq:th:1-3}
\lim_{t\rightarrow\tau^-}\|A(t,\tau)\|_p=\infty
\end{equation}
\end{thm}
\begin{pf}
Condition \eqref{eq:th:1-1} is a direct result of the upper bound of \eqref{eq:lem:1-1} in Lemma \ref{lem:1}. Given the limit assumption on the entries of $A(t,\tau)$, condition \eqref{eq:th:1-2} is obtainable from the lower bound of \eqref{eq:lem:1-1}. For condition \eqref{eq:th:1-3}, observe that according to the properties of logarithmic norm of a square matrix \cite{strom1975logarithmic}, $|\mu_p(A(t,\tau))|\leq\|A(t,\tau)\|_p$. Therefore, $|\mu_p(-A(t,\tau))|\leq\|A(t,\tau)\|_p$ and condition \eqref{eq:th:1-2} concludes \eqref{eq:th:1-3}.\hfill$\qed$
\end{pf}

\begin{cor}
\label{cor:1}
The matrix of a prescribed-time attractive LTV system, $A(t,\tau)$, has at least one entry with a diverging behavior at $t=\tau$, i.e., $\exists i,j\in\mathbb{N}:\lim_{t\rightarrow\tau^-}|A_{ij}(t,\tau)|=\infty$. Therefore:
\begin{enumerate}
\item Singularity at $t=\tau$ is necessary for prescribed-time attractive LTV systems. 
\item The closed-loop response of an LTI system $\dot{x}=Fx+Gu$ under a time-varying controller $u=K(t,\tau)x$ cannot be PTA unless the $p$-norm of the controller gain approaches infinity. A similar conclusion can be made for the gains of prescribed-time observers.
\end{enumerate}

\end{cor}

\begin{rem}
{Note that the necessary conditions \eqref{eq:th:1-2} and \eqref{eq:th:1-3} of Theorem \ref{th:1} are not valid if the limit conditions on the entries of $A(t,\tau)$ are violated. As a counter example, consider the scalar system $\dot{x}=a(t,\tau)x$ with $a(t,\tau)=-\left\{0.5-\sin[1/(\tau-t)]\right\}/(\tau-t)^3$ which its limit does not exist as $t\rightarrow\tau$. This system is PTA but its eigenvalue oscillates between $-\infty$ and $+\infty$.}
\end{rem}

\begin{rem}
\label{rem:2}
For nonscalar LTV systems, condition \eqref{eq:th:1-3} does not guarantee prescribed-time attractivity, even if the frozen-time eigenvalues have negative real parts (especially for scalar systems, a similar point is mentioned in \cite[Remark 1]{zhou2020finite2}). As a counter example, consider a system as \eqref{eq:1} with $A(t,\tau)=\mathrm{diag}([-1, -1/(\tau-t)])$. 
\end{rem}

\begin{rem}
\label{rem:3}
For symmetric systems $A(t,\tau)=A^T(t,\tau)$ (or more generally, systems that are symmetric close to $t=\tau$, i.e., $\lim_{t\rightarrow\tau^-}A(t,\tau)=\lim_{t\rightarrow\tau^-}A^T(t,\tau)$), condition \eqref{eq:th:1-1} of Theorem \ref{th:1} can be substituted by $\lim_{t\rightarrow\tau^-}\int_{0}^t\lambda_{\mathrm{max}}(A(s,\tau))ds=-\infty$.
\end{rem}

We need the following lemmas to prove Theorems \ref{th:2} and \ref{th:3}:

\begin{lem}\textbf{\textup{\cite{amato2009finite}}}
\label{lem:2}
Given $R\in\mathbb{R}^{n\times n}$ and $\Gamma(\cdot):[0,\infty)\rightarrow\mathbb{R}^{n\times n}$, condition
\begin{equation}
\label{eq:lem:2-1}
x^T(0)Rx(0)<1\Rightarrow x^T(t)\Gamma(t)x(t)<1
\end{equation}
is satisfied for an LTV system $\dot{x}=A(t)x$, if and only if the following differential Lyapunov
inequality, with terminal and initial conditions, has a piecewise continuously differentiable symmetric solution $P(t)$:
\begin{equation}
\label{eq:lem:2-2}
\dot{P}(s)+A^T(s)P(s)+P(s)A(s)<0,\hspace{2mm}s\in(0,t)
\end{equation}
\begin{equation}
\label{eq:lem:2-3}
P(t)\geq\Gamma(t)
\end{equation}
\begin{equation}
\label{eq:lem:2-4}
P(0)<R
\end{equation}
\end{lem}

\begin{lem}\textbf{\textup{\cite{fulton2000eigenvalues}}}
\label{lem:3}
Let $A$, $B$, and $C$ are $n\times n$ Hermitian matrices with eigenvalues $a_1\geq a_2\geq \cdots\geq a_n$, $b_1\geq b_2\geq \cdots\geq b_n$ and $c_1\geq c_2\geq \cdots\geq c_n$, respectively. If $A+B=C$, then:
\begin{equation}
\label{eq:lem:3-1}
a_i+b_j\leq c_k,\hspace{2mm}\mathrm{if} \hspace{2mm} i+j=n+k
\end{equation}
\end{lem}

\begin{lem}
\label{lem:4}
Let $A\in\mathbb{R}^{n\times n}$ and $B\in\mathbb{R}^{n\times n}$ are symmetric. If $A+B\leq0$ (respectively, $A+B<0$), then $\lambda_{\mathrm{min}}(A)+\lambda_{\mathrm{max}}(B)\leq0$ (respectively, $\lambda_{\mathrm{min}}(A)+\lambda_{\mathrm{max}}(B)<0$).
\end{lem}
\begin{pf}
Inequality $A+B\leq0$ (respectively, $A+B<0$) can be written as $A+B=C$ for some $C\leq0$ (respectively, $C<0$). The result is then straightforward by applying Lemma \ref{lem:3} with $i=n$ and $j=k=1$, and knowing that $\lambda_{\mathrm{max}}(C)\leq0$ (respectively, $\lambda_{\mathrm{max}}(C)<0$). \hfill$\qed$
\end{pf}

Lemma \ref{lem:2} gives necessary and sufficient conditions for finite-time stability of LTV systems such that the system solution is constrained by a time-varying hyper-ellipsoid. The following theorem makes use of Lemma \ref{lem:2} to formulate a condition equivalent to the prescribed-time attractivity of LTV systems.

\begin{thm}
\label{th:2}
System \eqref{eq:1} is PTA at $t=\tau$ if and only if there exists $\varepsilon>0$ such that the following conditions admit a continuously differentiable $P(t)$:
\begin{equation}
\label{eq:th:2-1}
\begin{split}
\dot{P}(t)+A^T(t,\tau)P(t)+P(t)A(t,\tau)<0,\ t\in[\tau-\varepsilon,\tau)
\end{split}
\end{equation}
\begin{equation}
\label{eq:th:2-2}
\lim_{t\rightarrow\tau^-}\lambda_{\mathrm{min}}(P(t))=\infty
\end{equation}
\end{thm}
\begin{pf}
Speaking in terms of Lemma \ref{lem:2}, the prescribed-time attractivity of an LTV system is equivalent to saying that the final hyper-ellipsoid of condition \eqref{eq:lem:2-1} becomes zero, i.e.,  $\lim_{t\rightarrow\tau^-}\lambda_{\mathrm{min}}(\Gamma(t))=\infty$. Besides, apply the rule of Lemma \ref{lem:4} to condition \eqref{eq:lem:2-3} to obtain $\lambda_{\mathrm{min}}(\Gamma(t))+\lambda_{\mathrm{max}}(-P(t))\leq0\Rightarrow\lambda_{\mathrm{min}}(\Gamma(t))\leq\lambda_{\mathrm{min}}(P(t))$. Thus, condition \eqref{eq:lem:2-3} needs the satisfaction of \eqref{eq:th:2-2}. As the prescribed-time attractivity is about the system's behavior when $t\rightarrow\tau$, there is no other restrictions on matrix $\Gamma(t)$, and condition \eqref{eq:lem:2-2} reduces to \eqref{eq:th:2-1}.\hfill$\qed$
\end{pf}

\begin{rem}
\label{rem:3}
Condition \eqref{eq:th:1-1} of Theorem \ref{th:1} is generally conservative, and its application in the design of PTA systems is limited to particular forms of systems (e.g., symmetric systems).  Also, the necessary and sufficient condition discussed in Theorem \ref{th:2} does not seem to facilitate the design process in its present form. The design of PTA systems can be accomplished in a more efficient way by other methods proposed in the literature \cite{song2017time,holloway2019prescribed,krishnamurthy2020dynamic,zhou2020finite,zhou2020finite2,zhou2021prescribed,shakouri2021framework}. 
\end{rem}

\begin{thm}
\label{th:3}
{Let system \eqref{eq:1} be PTA at $t=\tau$ such that $\lim_{t\rightarrow\tau-}|A_{i,j}(t,\tau)|$ exists or is infinity for all $i,j\in\mathbb{N}$.} Then, there exists $\varepsilon>0$ such that for all $t\in[\tau-\varepsilon,\tau)$ the frozen-time system is asymptotically stable, i.e.:
\begin{equation}
\label{eq:th:3-1}
\exists \varepsilon>0:\ \lambda_{\mathrm{max}}(A(t,\tau))<0,\ \forall t\in[\tau-\varepsilon,\tau)
\end{equation}
\end{thm}
\begin{pf}
{As the system is PTA, we know from Theorem \ref{th:2} that \eqref{eq:th:2-1} and \eqref{eq:th:2-2} hold true. Given condition \eqref{eq:th:2-1} and our assumption on the entries of $A(t,\tau)$, one can verify that $\lim_{t\rightarrow\tau-}\dot{P}_{i,j}(t)$ exists or is infinity for all $i,j\in\mathbb{N}$. Multiply both sides of $P(t)=\int_{t_0}^t\dot{P}(s)ds$ (which is valid for some $t_0$) by an arbitrary constant vector $v$ and its transpose, then verify that the following inequality holds:
\begin{equation}
\label{eq:th:3-2}
\lambda_{\mathrm{min}}(P(t))\|v\|^2\leq\left\|\int_{t_0}^tv^T\dot{P}(s)vds\right\|\leq \lambda_{\mathrm{max}}(P(t))\|v\|^2
\end{equation}
Given \eqref{eq:th:2-2} we know that $\lambda_{\mathrm{min}}(P(t))$ approaches infinity as $t\rightarrow\tau$. Thus, from \eqref{eq:th:3-2} is can be seen that $\lim_{t\rightarrow\tau^-}\left\|\int_{t_0}^tv^T\dot{P}(s)vds\right\|=\infty$. As the limit of $v^T\dot{P}(s)v$ exists or is infinity, it must satisfy $\lim_{t\rightarrow\tau^-}v^T\dot{P}(t)v=\infty$ for every $v$. Therefore, matrix $\dot{P}$ should eventually be positive-definite and satisfy $\lim_{t\rightarrow\tau^-}\lambda_{\mathrm{min}}(\dot{P}(t))=\infty$.} Apply the result of Lemma \ref{lem:4} to condition \eqref{eq:th:2-1} and obtain:
\begin{equation}
\label{eq:th:3-3}
\lambda_{\mathrm{max}}(A^T(t,\tau)P(t)+P(t)A(t,\tau))<-\lambda_{\mathrm{min}}(\dot{P}(t))
\end{equation}
Substitute $\lim_{t\rightarrow\tau^-}\lambda_{\mathrm{min}}(\dot{P}(t))=\infty$ into \eqref{eq:th:3-3} to obtain:
\begin{equation}
\label{eq:th:3-4}
\lim_{t\rightarrow\tau^-}\lambda_{\mathrm{max}}\left(A^T(t,\tau)P(t)+P(t)A(t,\tau)\right)=-\infty
\end{equation}
Observe that \eqref{eq:th:3-4} means that as $t$ approaches $\tau$, the term $A^T(t,\tau)P(t)+P(t)A(t,\tau)$ should be negative-definite, and therefore, matrix $A(t,\tau)$ should be Hurwitz at every $t$ sufficiently close to $\tau$.\hfill$\qed$
\end{pf}

\begin{cor}
\label{cor:2}
{Let system \eqref{eq:1} be PTA at $t=\tau$ such that $\lim_{t\rightarrow\tau-}|A_{i,j}(t,\tau)|$ exists or is infinity for all $i,j\in\mathbb{N}$.} Let $A_s(t,\tau,\sigma)$ be defined as follows:
\begin{equation}
\label{eq:cor:2-1}
A_s(t,\tau,\sigma)=\left\{\begin{array}{lcl}
A(t,\tau) & \hspace{2mm}\mathrm{if}\hspace{2mm} & \|x(t)\|>\sigma \\
A(t_s,\tau) & \hspace{2mm}\mathrm{if}\hspace{2mm} & \|x(t)\|\leq\sigma
\end{array}\right.
\end{equation}
where $t_s<\tau$ is the switching time. Then, there exists $\bar{\sigma}>0$ such that the solution of the switching system $\dot{x}=A_s(t,\tau,\sigma)x$ satisfies $\lim_{t\rightarrow\tau}\|x(t)\|\leq\sigma$ for every $\sigma\in(0,\bar{\sigma}]$ and the system is asymptotically stable at all $t=[t_s,\infty)$. Therefore, the answer to Problem \ref{prob:1} is positive. 
\end{cor}

\begin{cor}
\label{cor:3}
Suppose that $K(t,\tau)$ is a time-varying PTA state-feedback gain for the (unknown) LTI system $\dot{x}=Fx+Gu$ such that $\lim_{t\rightarrow\tau-}|K_{i,j}(t,\tau)|$ exists or is infinity for all $i,j\in\mathbb{N}$ and $u=K(t,\tau)x$ is guaranteed to be bounded. Let $K_s(t,\tau,\sigma)$ be defined as follows: 
\begin{equation}
\label{eq:cor:2-1}
K_s(t,\tau,\sigma)=\left\{\begin{array}{lcl}
K(t,\tau) & \hspace{2mm}\mathrm{if}\hspace{2mm} & \|x(t)\|>\sigma \\
K(t_s,\tau) & \hspace{2mm}\mathrm{if}\hspace{2mm} & \|x(t)\|\leq\sigma
\end{array}\right.
\end{equation}
where $t_s<\tau$ is the switching time. Then, there exists $\bar{\sigma}>0$ such that the solution of $\dot{x}=Fx+Gu$ under the control law $u=K_s(t,\tau,\sigma)x$ satisfies $\lim_{t\rightarrow\tau}\|x(t)\|\leq\sigma$ for every $\sigma\in(0,\bar{\sigma}]$, the constant state-feedback gain $K(t_s,\tau)$ asymptotically stabilizes the system, and $u=K_s(t,\tau,\sigma)x$ is bounded for all $t\in[0,\infty)$.
\end{cor}

\section{{Example}}
\label{sec:IV}

\begin{figure}[!h]
\centering\includegraphics[width=1\linewidth]{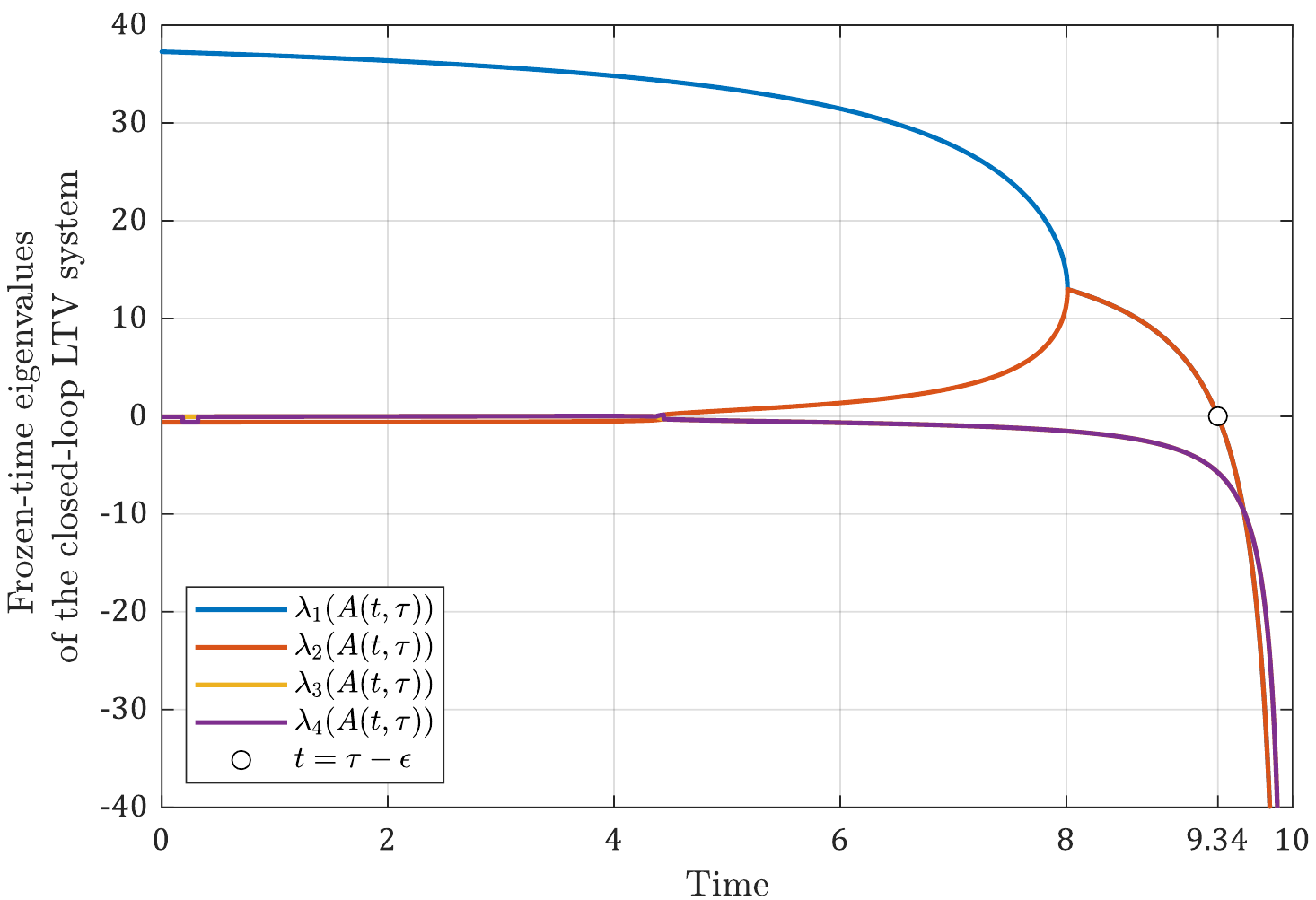}
\caption{{Frozen-time eigenvalues of an LTI system under a time-varying prescribed-time controller intended to converge at $\tau=10$.}}
\label{fig:1}
\end{figure}

In this section, we show the application of Theorem \ref{th:3} and Corollary \ref{cor:3} in the prescribed-time control of a typical single-input single-output (SISO) unknown LTI system. Consider system $\dot{x}=Fx+Gu$, expressed in controllable canonical form with unstable open-loop dynamics, defined as follows:
\begin{equation}
\label{eq:ex:1}
F=\begin{bmatrix}
0 & 1 & 0 & 0 \\
0 & 0 & 1 & 0 \\
0 & 0 & 0 & 1 \\
10 & 20 & 30 & 40
\end{bmatrix},\ G=\begin{bmatrix}
0 \\
0 \\
0 \\
1
\end{bmatrix}
\end{equation}
A prescribed-time controller $u=K(t,\tau)x$ can be selected for this system regardless of the model with a time-varying gain $K(t,\tau)=[K_{11},\ K_{12},\ K_{13},\ K_{14}]$ formulated as \cite[Example 2]{shakouri2021prescribed}\footnote{MATLAB\textsuperscript{\tiny\textregistered} codes and Simulink\textsuperscript{\tiny\textregistered} models for some prescribed-time controllers can be found in \href{https://github.com/a-shakouri/prescribed-time-control}{https://github.com/a-shakouri/prescribed-time-control}}:
\begin{subequations}
\label{eq:ex:2}
\begin{align}
K_{11}(t,\tau)&=-\frac{1}{\alpha^{4}(\tau-t)^{4}} \\
K_{12}(t,\tau)&=\frac{1}{(\tau-t)^{3}}\left(-\frac{4}{\alpha^{3}}+\frac{6}{\alpha^{2}}-\frac{4}{\alpha}+1\right)\\
K_{13}(t,\tau)&=\frac{1}{(\tau-t)^{2}}\left(-\frac{6}{\alpha^{2}}+\frac{12}{\alpha}-7\right) \\
K_{14}(t,\tau)&=\frac{1}{\tau-t}\left(-\frac{4}{\alpha}+6\right)
\end{align}
\end{subequations}
where $\alpha>0$ is a small constant. It can be seen from \eqref{eq:ex:2} that the entries of the controller gain approach infinity as $t\rightarrow\tau$, which is an obstacle in the practical implementation of $u=K(t,\tau)x$ (note that although the gain entries approach infinity, the control input remains bounded as proved in \cite{shakouri2021prescribed}). From Theorem \ref{th:3}, we know that the frozen-time eigenvalues of the closed-loop system $A(t,\tau)=F+GK(t,\tau)$ have negative real parts close to $t=\tau$. To illustrate this fact, we select $\alpha=0.1$, $\tau=10$, and $x(0)=[1,\ 1,\ 1,\ 1]^T$ for a simulation. Fig. \ref{fig:1} shows the frozen-time eigenvalues of $A(t,\tau)$, in which the eigenvalues have negative real parts at all $t\in[\tau-\varepsilon,\tau)=[9.34,10)$ where $\varepsilon=0.66$. For instance, $K(9.5,10)=-[1.6\times10^5,\ 2.7512\times10^4,\ 1.948\times10^3,\ 68]$ is a constant control gain with which system \eqref{eq:ex:1} is asymptotically stable. Therefore, a prescribed-time controller as proposed by Corollary \ref{cor:3}, $u=K_s(t,\tau,\sigma)x$, stabilizes the system at $t=\tau$ with an arbitrary error bound $\sigma>0$, while the unknown system is asymptotically stable after $t=\tau$, if $\sigma$ is selected small enough such that $\|x(t)\|\leq\sigma$ occurs after $t=\tau-\varepsilon$. 

\section{Conclusions}
\label{sec:V}

It has been proved that singularity at the convergence time is necessary for prescribed-time attractive (PTA) systems. Also, it has been proved that the real parts of the frozen-time eigenvalues, corresponding to a PTA system, become negative before the singularity moment. Therefore, by a state-dependent switching, a linear time-invariant system under a singular time-varying prescribed-time controller (or observer) reaches any nonzero error bound at the intended convergence time without numerical problems such that the closed-loop system is asymptotically stable after the switching occurs. 

\bibliography{autosam}

@article{bhat2000finite,
  title={Finite-time stability of continuous autonomous systems},
  author={Bhat, Sanjay P and Bernstein, Dennis S},
  journal={SIAM Journal on Control and Optimization},
  volume={38},
  number={3},
  pages={751--766},
  year={2000},
  publisher={SIAM},
  doi={10.1137/S0363012997321358}
}

@article{polyakov2011nonlinear,
  title={Nonlinear feedback design for fixed-time stabilization of linear control systems},
  author={Polyakov, Andrey},
  journal={IEEE Transactions on Automatic Control},
  volume={57},
  number={8},
  pages={2106--2110},
  year={2011},
  publisher={IEEE},
  doi={10.1109/TAC.2011.2179869}
}

@article{song2017time,
  title={Time-varying feedback for regulation of normal-form nonlinear systems in prescribed finite time},
  author={Song, Yongduan and Wang, Yujuan and Holloway, John and Krstic, Miroslav},
  journal={Automatica},
  volume={83},
  pages={243--251},
  year={2017},
  publisher={Elsevier},
  doi={10.1016/j.automatica.2017.06.008}
 }

@article{krishnamurthy2020dynamic,
  title={A dynamic high-gain design for prescribed-time regulation of nonlinear systems},
  author={Krishnamurthy, Prashanth and Khorrami, Farshad and Krstic, Miroslav},
  journal={Automatica},
  volume={115},
  pages={108860},
  year={2020},
  publisher={Elsevier},
  doi={10.1016/j.automatica.2020.108860}
}

@article{vrabel2019note,
  title={A note on uniform exponential stability of linear periodic time-varying systems},
  author={Vrabel, Robert},
  journal={IEEE Transactions on Automatic Control},
  volume={65},
  number={4},
  pages={1647--1651},
  year={2019},
  publisher={IEEE},
  doi={10.1109/TAC.2019.2927949}
}

@article{amato2009finite,
  title={Finite-time stability of linear time-varying systems with jumps},
  author={Amato, Francesco and Ambrosino, Roberto and Ariola, Marco and Cosentino, Carlo},
  journal={Automatica},
  volume={45},
  number={5},
  pages={1354--1358},
  year={2009},
  publisher={Elsevier},
  doi={10.1016/j.automatica.2008.12.016}
}

@article{fulton2000eigenvalues,
  title={Eigenvalues, invariant factors, highest weights, and {S}chubert calculus},
  author={Fulton, William},
  journal={Bulletin of the American Mathematical Society},
  volume={37},
  number={3},
  pages={209--249},
  year={2000},
  doi={10.1090/S0273-0979-00-00865-X}
}

@article{strom1975logarithmic,
  title={On logarithmic norms},
  author={Str{\"o}m, Torsten},
  journal={SIAM Journal on Numerical Analysis},
  volume={12},
  number={5},
  pages={741--753},
  year={1975},
  publisher={SIAM},
  doi={10.1137/0712055}
}

@article{zhou2021prescribed,
  title={Prescribed-time stabilization of a class of nonlinear systems by linear time-varying feedback},
  author={Zhou, Bin and Shi, Yang},
  journal={IEEE Transactions on Automatic Control},
  year={2021},
  publisher={IEEE},
  doi={10.1109/TAC.2021.3061645}
}

@article{zhou2020finite,
  title={Finite-time stabilization of linear systems by bounded linear time-varying feedback},
  author={Zhou, Bin},
  journal={Automatica},
  volume={113},
  pages={108760},
  year={2020},
  publisher={Elsevier},
  doi={10.1016/j.automatica.2019.108760}
}

@article{holloway2019prescribed,
  title={Prescribed-time observers for linear systems in observer canonical form},
  author={Holloway, John and Krstic, Miroslav},
  journal={IEEE Transactions on Automatic Control},
  volume={64},
  number={9},
  pages={3905--3912},
  year={2019},
  publisher={IEEE},
  doi={10.1109/TAC.2018.2890751}
}

@article{lopez2018finite,
  title={Finite-time and fixed-time observer design: Implicit {L}yapunov function approach},
  author={Lopez-Ramirez, Francisco and Polyakov, Andrey and Efimov, Denis and Perruquetti, Wilfrid},
  journal={Automatica},
  volume={87},
  pages={52--60},
  year={2018},
  publisher={Elsevier},
  doi={10.1016/j.automatica.2017.09.007}
}

@article{engel2002continuous,
  title={A continuous-time observer which converges in finite time},
  author={Engel, Robert and Kreisselmeier, Gerhard},
  journal={IEEE Transactions on Automatic Control},
  volume={47},
  number={7},
  pages={1202--1204},
  year={2002},
  publisher={IEEE},
  doi={10.1109/TAC.2002.800673}
}

@book{blondel2012open,
  title={Open problems in mathematical systems and control theory},
  author={Blondel, Vincent D and Sontag, Eduardo D and Vidyasagar, Mathukumalli and Willems, Jan C},
  year={2012},
  publisher={Springer Science \& Business Media}
}

@article{amato1993new,
  title={New sufficient conditions for the stability of slowly varying linear systems},
  author={Amato, Francesco and Celentano, Giovanni and Garofalo, Franco},
  journal={IEEE Transactions on Automatic Control},
  volume={38},
  number={9},
  pages={1409--1411},
  year={1993},
  publisher={IEEE},
  doi={10.1109/9.237657}
}

@article{zhou2016asymptotic,
  title={On asymptotic stability of linear time-varying systems},
  author={Zhou, Bin},
  journal={Automatica},
  volume={68},
  pages={266--276},
  year={2016},
  publisher={Elsevier},
  doi={10.1016/j.automatica.2015.12.030}
}

@article{wu1974note,
  title={A note on stability of linear time-varying systems},
  author={Wu, M},
  journal={IEEE Transactions on Automatic Control},
  volume={19},
  number={2},
  pages={162--162},
  year={1974},
  publisher={IEEE},
  doi={10.1109/TAC.1974.1100529}
}

@article{zhou2020finite2,
  title={Finite-time stability analysis and stabilization by bounded linear time-varying feedback},
  author={Zhou, Bin},
  journal={Automatica},
  volume={121},
  pages={109191},
  year={2020},
  publisher={Elsevier},
  doi={10.1016/j.automatica.2020.109191}
}

@article{shakouri2021prescribed,
  title={Prescribed-time control with linear decay for nonlinear systems},
  author={Shakouri, Amir and Assadian, Nima},
  journal={IEEE Control Systems Letters},
  year={2022},
  volume={6},
  number={},
  pages={313-318},
  publisher={IEEE},
  doi={10.1109/LCSYS.2021.3073346}
}

@article{holloway2019prescribed2,
  title={Prescribed-time output feedback for linear systems in controllable canonical form},
  author={Holloway, John and Krstic, Miroslav},
  journal={Automatica},
  volume={107},
  pages={77--85},
  year={2019},
  publisher={Elsevier},
  doi={10.1016/j.automatica.2019.05.027}
}

@article{shakouri2021euler,
  title={Prescribed-time Control for Perturbed {E}uler-{L}agrange Systems with Obstacle Avoidance},
  author={Shakouri, Amir and Assadian, Nima},
  journal={IEEE Transactions on Automatic Control},
  year={2021},
  volume={},
  number={},
  pages={},
  publisher={IEEE},
  doi={10.1109/tac.2021.3106882}
}

@article{ding2020nonsingular,
  title={Nonsingular prescribed-time stabilization of a class of uncertain nonlinear systems: A novel coordinate mapping method},
  author={Ding, Chao and Shi, Chao and Chen, Yong},
  journal={International Journal of Robust and Nonlinear Control},
  volume={30},
  number={9},
  pages={3566--3581},
  year={2020},
  publisher={Wiley Online Library},
  doi={10.1002/rnc.4949}
}

@article{shakouri2021framework,
  title={A Framework for Prescribed-Time Control Design via Time-Scale Transformation },
  author={Shakouri, Amir and Assadian, Nima},
  journal={IEEE Control Systems Letters},
  year={2021},
  publisher={IEEE}
}

\end{document}